\documentstyle[aps,epsf,floats,twocolumn,prl]{revtex}
\begin{document}
\newcommand{\bgm}[1]{\mbox{\boldmath $#1$}}
\newcommand{\bgms}[1]{\mbox{{\scriptsize \boldmath $#1$}}}
\newcommand{\ul}[1]{\underline{#1}} \newcommand{\bgt}[1]{{\boldmath
    $#1$}}

\twocolumn[\hsize\textwidth\columnwidth\hsize\csname@twocolumnfalse%
\endcsname \draft

\title{Complexity Through Nonextensivity}

\author{William Bialek$^1$, Ilya Nemenman$^1$, and Naftali
  Tishby$^{1,2}$}

\address{$^1$NEC Research Institute, 4 Independence Way,
  Princeton, New Jersey 08540\\
  $^2$School of Computer Science and Engineering,\\
  and Center for Neural Computation, Hebrew University, Jerusalem
  91904, Israel}

\date{\today} \maketitle

\begin{abstract}
  The problem of defining and studying complexity of a time series has
  interested people for years. In the context of dynamical systems,
  Grassberger has suggested that a slow approach of the entropy to its
  extensive asymptotic limit is a sign of complexity. We investigate
  this idea further by information theoretic and statistical mechanics
  techniques and show that these arguments can be made precise, and
  that they generalize many previous approaches to complexity, in
  particular unifying ideas from the physics literature with ideas
  from learning and coding theory; there are even connections of this
  statistical approach to algorithmic or Kolmogorov complexity.
  Moreover, a set of simple axioms similar to those used by Shannon in
  his development of information theory allows us to prove that the
  divergent part of the subextensive component of the entropy is a
  unique complexity measure. We classify time series by their
  complexities and demonstrate that beyond the `logarithmic'
  complexity classes widely anticipated in the literature there are
  qualitatively more complex, `power--law' classes which deserve more
  attention.

\end{abstract}
\pacs{PACS} ]

The problem of quantifying complexity is very old. Interest in the
field has been fueled by three sorts of questions.  First, one would
like to make precise an impression that some systems, such as life on
earth or a turbulent fluid flow, evolve toward a state of higher
complexity, and one would like to classify these states; this is the
realm of dynamical systems theory.  Second, in choosing among
different models that describe an experiment, one wants to quantify a
preference for simpler explanations or, equivalently, provide a
penalty for complex models that can be weighed against the more
conventional goodness of fit criteria; this type of question usually
is investigated in statistics. Finally, there are questions about how
hard it is to compute or to describe the state of a complex system;
this is the area of formal mathematics and computer science.

Research in each of these three directions has given birth to numerous
definitions of complexity. The usual objective is to make these
definitions focused enough to be operational in particular contexts
but general enough to connect with our intuitive notions.  For many
years the dominant candidate for a universal measure has been the
mathematically rigorous notion of {\em Kolmogorov} or {\em algorithmic
  complexity} that measures (roughly) the minimum length of a computer
program that can recreate the observed time series \cite{lv_book}.
Unfortunately there is no algorithm that can calculate the Kolmogorov
complexity of all data sets. Therefore, for applications to
statistics, Rissanen \cite{riss_book} and others have developed a new
concept: {\em stochastic complexity} of the data with respect to a
particular class of models, which measures the shortest total
description of the data and the model within the class, but cannot
rule out the possibility that a different model class could generate a
shorter code.

The main difficulty of all these approaches is that the Kolmogorov
complexity is closely related to the Shannon entropy, which means that
it measures something closer to our intuitive concept of randomness
than to the intuitive concept of complexity \cite{bennet90}. A true
random string cannot be compressed and hence requires a long
description, yet the physical process that generates this string may
be very simple. As physicists, our intuitive notions of complexity
correspond to statements about the underlying process, and not
directly to the description length or Kolmogorov complexity: a
dynamics with a predictable constant output (small algorithmic
complexity) is as trivial as one for which the output is completely
unpredictable and random (large algorithmic complexity), while really
complex processes lie somewhere in between.

The two extreme cases, however, have one feature in common: the
entropy of the output strings (or, equivalently, the Kolmogorov
complexity of a typical one) either is a fixed constant or grows
exactly linearly with the length of the strings. In both cases,
corrections to the asymptotic behavior {\em do not grow with the size
  of the data set}.  This allowed Grassberger \cite{grassberger} to
identify the slow approach of the entropy to its extensive limit as a
sign of complexity. He has proposed several functions to analyze this
slow approach and studied systems that exhibited a broad range of
complexity properties.

To deal with the same problem, Rissanen has emphasized strongly that
fitting a model to data represents an encoding of those data, or
predicting future data. Shorter encodings generally mean better
prediction or generalization. However, much of the code usually
describes the meaningless, nongeneralizable ``noise''---statistical
fluctuations within the model.  Only model description is relevant to
prediction, and this part of the code has been termed the {\em model
  complexity} \cite{riss_book}.  While systems with model complexity
of very different types are known, the two extreme examples above are
similar: it only takes a fixed number of bits to code either a call to
a random number generator or to a constant function.

The present work may be viewed as expanding on the notions of
subextensivity and effective prediction. We construct a coherent
theory that brings these ideas together in an intuitive way, but
nonetheless is sufficiently general to be applied in many different
contexts. We will show that with only a little bit of work
Grassberger's definitions may be made as mathematically precise as
they are aesthetically pleasing. Finally, we will argue that the
definitions are unique if one accepts a set of simple axioms in the
spirit of Shannon's original work, and that these definitions relate
to the usual Kolmogorov complexity in a straightforward way.  Much of
this paper follows closely a more detailed analysis in Ref.\ 
\cite{bnt}, to which we refer for calculation details and a thorough
discussion of the relevant literature.

Our path to connecting the various complexity measures begins by
noticing that the subextensive components of entropy identified by
Grassberger in fact determine the information available for making
predictions.  This also suggests a connection to the importance or
value of information, especially in a biological or economic context:
information is valuable if it can be used to guide our actions, but
actions take time and hence observed data can be useful only to the
extent that those data inform us about the state of the world at later
times.  It would be attractive if what we identify as ``complex'' in a
time series were also the ``useful'' or ``meaningful'' components.

While prediction may come in various forms, depending on context,
information theory allows us to treat all of them on the same footing.
For this we only need to recognize that all predictions are
probabilistic, and that, even before we look at the data, we know that
certain futures are more likely than others. This knowledge can be
summarized by a prior probability distribution for the futures.  Our
observations on the past lead us to a new, more tightly concentrated
distribution, the distribution of futures conditional on the past
data. Different kinds of predictions are different slices through or
averages over this conditional distribution, but information theory
quantifies the ``concentration'' of the distribution without making
any commitment as to which averages will be most interesting.

Imagine that we observe a stream of data $x(t)$ over a time interval
$-T < t < 0$; let all of these past data be denoted by the shorthand
$x_{\rm past}$.  We are interested in saying something about the
future, so we want to know about the data $x(t)$ that will be observed
in the time interval $0 < t < T'$; let these future data be called
$x_{\rm future}$.  In the absence of any other knowledge, futures are
drawn from the probability distribution $P(x_{\rm future})$, while
observations of particular past data $x_{\rm past}$ tell us that
futures will be drawn from the conditional distribution $P(x_{\rm
  future} | x_{\rm past})$. The greater concentration of the
conditional distribution can be quantified by the fact that it has
smaller entropy than the prior distribution, and this reduction in
entropy is Shannon's definition of the information that the past
provides about the future.  We can write the average of this {\em
  predictive information} as
\begin{eqnarray}
{\cal I}_{\rm pred} (T,T') &=& 
{\Bigg\langle} \log_2 \left[ {{P(x_{\rm future}| x_{\rm past})} 
\over{P(x_{\rm future})}}\right]\Bigg\rangle
  \\ 
&=& -\langle\log_2 P(x_{\rm future})\rangle 
- \langle\log_2 P( x_{\rm past})\rangle
\nonumber\\
&&\,\,\,\,\,\,\,\,\,\, 
-\left[-\langle\log_2 P(x_{\rm future}, x_{\rm past})\rangle\right]\,,
\label{ents}
\end{eqnarray}
where $\langle \cdots \rangle$ denotes an average over the joint
distribution of the past and the future, $P(x_{\rm future} , x_{\rm
  past})$.

Each of the terms in Eq.~(\ref{ents}) is an entropy. Since we are
interested in predictability or generalization, which are associated
with some features of the signal persisting forever, we may assume
stationarity or invariance under time translations. Then the entropy
of the past data depends only on the duration of our observations, so
we can write $ -\langle\log_2 P( x_{\rm past})\rangle = S(T) $, and by
the same argument $-\langle\log_2 P( x_{\rm future})\rangle = S(T')$.
Finally, the entropy of the past and the future taken together is the
entropy of observations on a window of duration $T+T'$, so that $
-\langle\log_2 P(x_{\rm future} , x_{\rm past})\rangle = S(T+T')$.
Putting these equations together, we obtain
\begin{equation}
{\cal I}_{\rm pred}(T,T') = S(T) +S(T') - S(T+T') . \label{IpredandST}
\end{equation}

In the same way that the entropy of a gas at fixed density is
proportional to the volume, the entropy of a time series
(asymptotically) is proportional to its duration, so that
$\lim_{T\rightarrow\infty} {{S(T)}/ T} = {\cal S}_0$; entropy is an
extensive quantity.  But from Eq.~(\ref{IpredandST}) any extensive
component of the entropy cancels in the computation of the predictive
information: {\em predictability is a deviation from extensivity}.  If
we write
\begin{equation}
  S(T) = {\cal S}_0 T +S_1(T)\,,
\end{equation}
then Eq.~(\ref{IpredandST}) tells us that the predictive information
is related {\em only} to the nonextensive term $S_1(T)$.

We know two general facts about the behavior of $S_1(T)$.  First, the
corrections to extensive behavior are positive, $S_1(T) \geq 0$.
Second, the statement that entropy is extensive is the statement that
the limit $ \lim_{T\rightarrow\infty} {{S(T)} / T} = {\cal S}_0 $
exists, and for this to be true we must also have $ \lim_{ T
  \rightarrow \infty} {{S_1(T)} / T} = 0.$ Thus the nonextensive terms
in the entropy must be {\em sub}extensive, that is they must grow with
$T$ less rapidly than a linear function.  Taken together, these facts
guarantee that the predictive information is positive and
subextensive.  Further, if we let the future extend forward for a very
long time, $T' \rightarrow \infty$, then we can measure the
information that our sample provides about the entire future,
\begin{equation}
 I_{\rm pred} (T) = 
\lim_{T' \rightarrow \infty} {\cal I}_{\rm pred}(T,T') = S_1 (T)\,,
\end{equation}
and this is precisely equal to the subextensive entropy.

If we have been observing a time series for a (long) time $T$, then
the total amount of data we have collected in is measured by the
entropy $S(T)$, and at large $T$ this is given approximately by ${\cal
  S}_0 T$.  But the predictive information that we have gathered
cannot grow linearly with time, even if we are making predictions
about a future which stretches out to infinity. As a result, of the
total information we have taken in by observing $x_{\rm past}$, only a
vanishing fraction is of relevance to the prediction:
\begin{equation}
\lim_{T\rightarrow\infty} {{\rm Predictive\ Information} \over{\rm
Total\ Information}} = {{I_{\rm pred} (T)} \over {S(T)}}
\rightarrow 0. \label{chuck}
\end{equation}
In this precise sense, most of what we observe is irrelevant to the
problem of predicting the future.  Since the average Kolmogorov
complexity of a time series is related to its (total) Shannon entropy,
this result means also that most of the algorithm that is required to
encode the data encodes aspects of the data that are useless for
prediction or for guiding our actions based on the data.  This is a
strong indication that the usual notions of Kolmogorov complexity in
fact do not capture anything at all like the (intuitive) utility of
the data stream.

Consider the case where time is measured in discrete steps, so that we
have seen $N$ time points $x_1, x_2 , \cdots , x_N$. How much is there
to learn about the underlying pattern in these data? In the limit of
large number of observations, $ N \to \infty$ or $T \to \infty$ the
answer to this question is surprisingly universal: predictive
information may either stay finite, or grow to infinity together with
$T$; in the latter case the rate of growth may be slow (logarithmic)
or fast (sublinear power).

The first possibility, $\lim_{T\rightarrow\infty} I_{\rm pred} (T) = $
constant, means that no matter how long we observe we gain only a
finite amount of information about the future. This situation
prevails, in both extreme cases mentioned above. For example, when the
dynamics are too regular, such as it is for a purely periodic system,
complete prediction is possible once we know the phase, and if we
sample the data at discrete times this is a finite amount of
information; longer period orbits intuitively are more complex and
also have larger $I_{\rm pred}$, but this doesn't change the limiting
behavior $\lim_{T\rightarrow\infty} I_{\rm pred} (T) =$ constant.

Similarly, the predictive information can be small when the dynamics
are irregular but the best predictions are controlled only by the
immediate past, so that the correlation times of the observable data
are finite. This happens, for example, in many physical systems far
away from phase transitions. Imagine, for example, that we observe
$x(t)$ at a series of discrete times $\{t_{\rm n}\}$, and that at each
time point we find the value $x_{\rm n}$. Then we always can write the
joint distribution of the $N$ data points as a product,
\begin{eqnarray}
P(x_1 , x_2 , \cdots , x_N ) &=& P(x_1 ) P(x_2 | x_1) 
P(x_3 | x_2 , x_1) \cdots . 
\end{eqnarray}
For Markov processes, what we observe at $t_{\rm n}$ depends only on
events at the previous time step $t_{\rm n-1}$, so that
\begin{eqnarray}
P(x_{\rm n} | \{x_{\rm 1\leq i \leq n-1}\}) &=& 
P(x_{\rm n} | x_{\rm n-1}) ,
\end{eqnarray}
and hence the predictive information reduces to
\begin{eqnarray}
I_{\rm pred} = \Bigg\langle \log_2 \left[ {{P(x_{\rm n}|x_{\rm n-1})}
\over{P(x_{\rm n})}} \right] \Bigg\rangle .
\end{eqnarray}
The maximum possible predictive information in this case is the
entropy of the distribution of states at one time step, which in turn
is bounded by the logarithm of the number of accessible states. To
approach this bound the system must maintain memory for a long time,
since the predictive information is reduced by the entropy of the
transition probabilities. Thus systems with more states and longer
memories have larger values of $I_{\rm pred}$.

More interesting are those cases in which $I_{\rm pred}(T)$ diverges
at large $T$. In physical systems we know that there are critical
points where correlation times become infinite, so that optimal
predictions will be influenced by events in the arbitrarily distant
past. Under these conditions the predictive information can grow
without bound as $T$ becomes large; for many systems the divergence is
logarithmic, $I_{\rm pred} (T\rightarrow\infty) \propto \log T$.

Long range correlation also are important in a time series where we
can learn some underlying rules. Suppose a series of random vector
variables $\{{\vec x}_{\rm i}\}$ are drawn independently from the same
probability distribution $Q({\vec x} | {\bgm\alpha})$, and this
distribution depends on a (potentially infinite dimensional) vector of
parameters $\bgm{\alpha}$. The parameters are unknown, and before the
series starts they are chosen randomly from a distribution ${\cal
  P}(\bgm{\alpha})$.  In this setting, at least implicitly, our
observations of $\{{\vec x}_{\rm i}\}$ provide data from which we can
learn the parameters $\bgm{\alpha}$.  Here we put aside (for the
moment) the usual problem of learning---which might involve
constructing some estimation or regression scheme that determines a
``best fit'' $\bgm{\alpha}$ from the data $\{{\vec x}_{\rm i}\}$---and
treat the ensemble of data streams $P[\{{\vec x}_{\rm i}\}]$ as we
would any other set of configurations in statistical mechanics or
dynamical systems theory.  In particular, we can compute the entropy
of the distribution $P[\{{\vec x}_{\rm i}\}]$ even if we can't provide
explicit algorithms for solving the learning problem.

As is shown in \cite{bnt}, the crucial quantity in such analysis is
the density of models in the vicinity of the target
$\bar{\bgm\alpha}$---the parameters that actually generated the
sequence. For two distributions, a natural distance measure is the
Kullback--Leibler divergence $D(\bar{\bgm\alpha} || {\bgm\alpha}) =
\int d{\vec x} Q({\vec x} | \bar{\bgm\alpha})\log \left[ Q({\vec x} |
  \bar{\bgm\alpha}) /Q({\vec x} | {\bgm\alpha}) \right]$, and the
density is
\begin{equation}
\rho (D;{\bar{\bgm\alpha}}) = \int d^K\alpha  {\cal P} ({\bgm\alpha})
\delta [D - D_{\rm KL} ( \bar{\bgm\alpha} || {\bgm\alpha} )].
\end{equation}
If $\rho$ is large as $D\to 0$, then one easily can get close to the
target for many different data; thus they are not very informative. On
the other hand, small density means that only very particular data
lead to $\bar{\bgm\alpha}$, so they carry a lot of predictive
information.  Therefore, it is clear that the density, but not the
number of parameters or any other simplistic measure, characterizes
predictability and the complexity of prediction. If, as often is the
case for $\dim {\bgm\alpha} <\infty$, the density behaves in the way
common to finite dimensional systems of the usual statistical
mechanics,
\begin{equation}
  \rho(D\to 0, \bar{\bgm\alpha}) \approx A D^{(K-2)/2}\,,
\end{equation}
then the predictive information to the leading order is
\begin{equation}
  I_{\rm pred}(N) \approx K/2\, \log N\,.
\end{equation}

The modern theory of learning is concerned in large part with
quantifying the complexity of a model class, and in particular with
replacing a simple count of parameters with a more rigorous notion of
dimensionality for the space of models; for a general review of these
ideas see Ref.~\cite{vapnik}, and for discussion close in spirit to
ours see Ref.~\cite{vijay}.  The important point here is that the
dimensionality of the model class, and hence the complexity of the
class in the sense of learning theory, emerges as the coefficient of
the logarithmic divergence in $I_{\rm pred}$.  Thus a measure of
complexity in learning problems can be derived from a more general
dynamical systems or statistical mechanics point of view, treating the
data in the learning problem as a time series or one dimensional
lattice.  The logarithmic complexity class that we identify as being
associated with finite dimensional models also arises, for example, at
the Feigenbaum accumulation point in the period doubling route to
chaos \cite{grassberger}.

As noted by Grassberger in his original discussion, there are time
series for which the divergence of $I_{\rm pred}$ is stronger than a
logarithm.  We can construct an example by looking at the density
function $\rho$ in our learning problem above: finite dimensional
models are associated with algebraic decay of the density as
$D\rightarrow 0$, and we can imagine that there are model classes in
which this decay is more rapid, for example
\begin{equation}
  \rho (D \rightarrow 0)
  \approx A \exp\left[ -{B /{D^\mu}}\right] ,\,\ \mu>0 \,.
\end{equation}  
In this case it can be shown that the predictive information diverges
very rapidly, as a sublinear power law,
\begin{equation}
  I_{\rm pred} (N) \sim N^{\mu/(\mu+1)} \,.
\end{equation}
One way that this scenario can arise is if the distribution $Q({\vec
  x})$ that we are trying to learn does not belong to any finite
parameter family, but is itself drawn from a distribution that
enforces a degree of smoothness \cite{bcs}.  Understandably, stronger
smoothness constraints have smaller powers (less to predict) than the
weaker ones (more to predict).  For example, a rather simple case of
predicting a one dimensional variable that comes from a continuous
distribution produces $I_{\rm pred}(N) \sim \sqrt{N}$.

As with the logarithmic class, we expect that power--law divergences
in $I_{\rm pred}$ are not restricted to the learning problems that we
have studied in detail.  The general point is that such behavior will
be seen in problems where predictability over long scales, rather then
being controlled by a fixed set of ever more precisely known
parameters, is governed by a progressively more detailed
description---effectively increasing the number of parameters---as we
collect more data.  This seems a plausible description of what happens
in language, where rules of spelling allow us to predict forthcoming
letters of long words, grammar binds the words together, and
compositional unity of the entire text allows to make predictions
about the subject of the last page of the book after reading only the
first few.  Indeed, Shannon's classic experiment on the predictability
of English text (by human readers!) shows this behavior
\cite{shannon-lang}, and more recently several groups have extracted
power--law subextensive components from the numerical analysis of
large corpora of text (see, for example, \cite{ebpol}, \cite{gs}).

Interestingly, even without an explicit example, a simple argument
ensures existence of exponential densities and, therefore, power law
predictive information models.  If the number of parameters in a
learning problem is not finite then in principle it is impossible to
predict anything unless there is some appropriate regularization.  If
we let the number of parameters stay finite but become large, then
there is {\em more} to be learned and correspondingly the predictive
information grows in proportion to this number. On the other hand, if
the number of parameters becomes infinite without regularization, then
the predictive information should go to zero since nothing can be
learned.  We should be able to see this happen in a regularized
problem as the regularization weakens: eventually the regularization
would be insufficient and the predictive information would vanish.
The only way this can happen is if the predictive information grows
more and more rapidly with $N$ as we weaken the regularization, until
finally it becomes extensive (equivalently, drops to zero) at the
point where prediction becomes impossible. To realize this scenario we
have to go beyond $I_{\rm pred} \propto \log T$ with $I_{\rm pred}
\propto N^{\mu/(\mu+1)}$; the transition from increasing predictive
information to zero occurs as $\mu \to 1$.

This discussion makes it clear that the predictive information (the
subextensive entropy) distinguishes between problems of intuitively
different complexity and thus, in accord to Grassberger's definitions
\cite{grassberger}, is probably a good choice for a universal
complexity measure. Can this intuition be made more precise?

First we need to decide whether we want to attach measures of
complexity to a particular signal $x(t)$ or whether we are interested
in measures that are defined by an average over the ensemble
$P[x(t)]$. One problem in assigning complexity to single realizations
is that there can be atypical data streams. Second, Grassberger
\cite{grassberger} in particular has argued that our visual intuition
about the complexity of spatial patterns is an ensemble concept, even
if the ensemble is only implicit. The fact that we admit probabilistic
models is crucial: even at a colloquial level, if we allow for
probabilistic models then there is a simple description for a sequence
of truly random bits, but if we insist on a deterministic model then
it may be very complicated to generate precisely the observed string
of bits. Furthermore, in the context of probabilistic models it hardly
makes sense to ask for a dynamics that generates a particular data
stream; we must ask for dynamics that generate the data with
reasonable probability, which is more or less equivalent to asking
that the given string be a typical member of the ensemble generated by
the model.  All of these paths lead us to thinking not about single
strings but about ensembles in the tradition of statistical mechanics,
and so we shall search for measures of complexity that are averages
over the distribution $P[x(t)]$.

Once we focus on average quantities, we can provide an axiomatic proof
(much in the spirit of Shannon's \cite{shannon} arguments establishing
entropy as a unique information measure) that links $I_{\rm pred}$ to
complexity. We can start by adopting Shannon's postulates as
constraints on a measure of complexity: if there are $N$ equally
likely signals, then the measure should be monotonic in $N$; if the
signal is decomposable into statistically independent parts then the
measure should be additive with respect to this decomposition; and if
the signal can be described as a leaf on a tree of statistically
independent decisions then the measure should be a weighted sum of the
measures at each branching point. We believe that these constraints
are as plausible for complexity measures as for information measures,
and it is well known from Shannon's original work that this set of
constraints leaves the entropy as the only possibility.  Since we are
discussing a time dependent signal, this entropy depends on the
duration of our sample, $S(T)$.  We know of course that this cannot be
the end of the discussion, because we need to distinguish between
randomness (entropy) and complexity.  The path to this distinction is
to introduce other constraints on our measure.

First we notice that if the signal $x$ is continuous, then the entropy
is not invariant under transformations of $x$ that do not mix point at
different times (reparameterizations).  It seems reasonable to ask
that complexity be a function of the process we are observing and not
of the coordinate system in which we choose to record our
observations. However, that it is not the whole function $S(T)$ which
depends on the coordinate system for $x$; it is only the extensive
component of the entropy that has this noninvariance.  This can be
seen more generally by noting that subextensive terms in the entropy
contribute to the mutual information among different segments of the
data stream (including the predictive information defined here), while
the extensive entropy cannot; mutual information is coordinate
invariant, so all of the noninvariance must reside in the extensive
term.  Thus, any measure complexity that is coordinate invariant must
discard the extensive component of the entropy.

If we continue along these lines, we can think about the asymptotic
expansion of the entropy at large $T$.  The extensive term is the
first term in this series, and we have seen that it must be discarded.
What about the other terms?  In the context of predicting in a
parameterized model, most of the terms in this series depend in detail
on our prior distribution in parameter space, which might seem odd for
a measure of complexity.  More generally, if we consider
transformations of the data stream $x(t)$ that mix points within a
temporal window of size $\tau$, then for $T >> \tau$ the entropy
$S(T)$ may have subextensive terms which are constant, and these are
not invariant under this class of transformations.  On the other hand,
if there are divergent subextensive terms, these {\em are} invariant
under such temporally local transformations \cite{note4}. So if we
insist that measures of complexity be invariant not only under
instantaneous coordinate transformations, but also under temporally
local transformations, then we can discard both the extensive and the
finite subextensive terms in the entropy, leaving only the divergent
subextensive terms as a possible measure of complexity.

To illustrate the purpose of these two extra conditions, we may think
of the following example: measuring velocity of a turbulent fluid flow
at a given point. The condition of invariance under
reparameterizations means that the complexity is independent of the
scale used by the speedometer. On the other hand, the second condition
ensures that the temporal mixing due to the finiteness of the inertia
of the speedometer's needle does not change the estimated complexity
of the flow.

In our view, these arguments (or their slight variation also presented
in \cite{bnt}) settle the question of the unique definition of
complexity. Not only is the divergent subextensive component of the
entropy the unique complexity measure, but it is also a universal one
since it is connected in a straightforward way to many other measures
that have arisen in statistics and in dynamical systems theory. A bit
less straightforward is the connection to the Kolmogorov's definition
that started the whole discussion, but even this can also be made.

To make this connection we follow the suggestion of Standish
\cite{standish} that one should focus not on the complexity of
particular strings but of equivalence classes.  In the present case it
is natural to define an equivalence class of data $x(-T < t \leq 0)$
as those data that generate indistinguishable conditional probability
distributions for the future, $P[x(t>0)\,|\,x(-T < t \leq 0)]$.  If
this conditional distribution has sufficient statistics, then there
exists a compression of the past data $x(-T < t \leq 0)$ into exactly
$I_{\rm pred}(T)$ bits while preserving all of the mutual information
with the future.  But this means that the ensemble of data in an
equivalence class can be described, on average, using exactly this
many bits.  Thus, for dynamics such that the prediction problem has
sufficient statistics, the average Kolmogorov complexity of
equivalence classes defined by the indistinguishability of predictions
is equal to the predictive information.  By the arguments above,
prediction is {\em the} useful thing which we can do with a data
stream, and so in this case it makes sense to say that the Kolmogorov
complexity of representing the useful bits of data is equal to the
predictive information. Note also that Kolmogorov complexity is
defined only up to a constant depending on the computer used
\cite{lv_book}. A computer independent definition requires ignoring
constant terms and focusing only on asymptotic behavior. This agrees
very well with our arguments above that identified only the divergent
part of the predictive information with the complexity of a data
stream.

In the terminology suggested by Grassberger, the statement that the
prediction problem has sufficient statistics means that the {\em True
  Measure Complexity} is equal to the {\em Effective Measure
  Complexity} \cite{grassberger}; similarly, the {\em statistical
  complexity} defined by Crutchfield and coworkers \cite{crutchfield}
then also is equal to predictive information defined here.  These are
strong statements, and it is likely that they are not true precisely
for most natural data streams.  More generally one can ask for
compressions that preserve the maximum fraction of the relevant (in
this case, predictive) information, and our intuitive notion of data
being ``understandable'' or ``summarizable'' is that these selective
compressions can be very efficient \cite{tpb}---here efficiency means
that we can compress the past into a description with length not much
larger than $I_{\rm pred}(T)$ while preserving a finite fraction of
the (diverging) information about the future; an example is when we
summarize data by the parameters of the model that describes the
underlying stochastic process. The opposite situation is illustrated
by certain cryptographic codes, where the relevant information is
accessible (at best) only from the entire data set.  Thus we can
classify data streams by their predictive information, but
additionally by whether this predictive information can be represented
efficiently.  For those data where efficient representation is
possible, the predictive information and the mean Kolmogorov
complexity of future--equivalent classes will be similar; with more
care we can guarantee that these quantities are proportional as
$T\rightarrow\infty$.  Perhaps Wigner's famous remarks about the
unreasonable effectiveness of mathematics in the natural sciences
could be rephrased as the conjecture that the data streams occurring
in nature---although often complex as measured by their predictive
information---nonetheless belong to this efficiently representable
class.


\begin{thebibliography}{99}
  
    \bibitem{lv_book}\newblock{M.\ Li and P.\ Vit{\'a}nyi. {\em An
      Introduction to Kolmogorov Complexity and its Applications},
    Springer--Verlag, New York (1993).}
  
    \bibitem{riss_book}\newblock{J.~Rissanen. {\em Stochastic
      Complexity and Statistical Inquiry}, World Scientific, Singapore
    (1989); J.~Rissanen, {\em IEEE Trans.\ Inf.\ Thy.} {\bf 42},
    40--47 (1996).}
  
    \bibitem{bennet90}\newblock{C.~Bennett, in {\it Complexity,
      Entropy and the Physics of Information}, W.~H.~Zurek, ed.,
    Addison--Wesley, Redwood City, pp.~137--148 (1990).}
  
    \bibitem{grassberger}\newblock{P.~Grassberger, {\it
      Int.~J.~Theor.~Phys.} {\bf 25}, 907--938 (1986).}
  
    \bibitem{bnt}\newblock{W.~Bialek, I.~Nemenman, and N.~Tishby, to
    appear in {\em Neural Computation} (2001). E-print: {\small \tt
      physics/0007070}.}
  
    \bibitem{vapnik}\newblock{V.~Vapnik. \emph{Statistical Learning
      Theory}, John Wiley \& Sons, New York (1998).}
  
    \bibitem{vijay}\newblock{V.~Balasubramanian, {\em Neural Comp.}
    {\bf 9}, 349--368 (1997).}
  
    \bibitem{bcs}\newblock{W.~Bialek, C.~Callan, and S.~Strong, {\it
      Phys.~Rev.~Lett.}  {\bf 77}, 4693--4697 (1996).}
  
    \bibitem{shannon-lang}\newblock{C.~E.~Shannon, {\em Bell Sys.\ 
      Tech.\ J.} {\bf 30}, 50--64 (1951). W.~Hilberg, {\it Frequenz}
    {\bf 44}, 243--248(1990).}
  
    \bibitem{ebpol}\newblock{W.~Ebeling, T.~P{\"o}schel, {\em
      Europhys.~Lett.}  {\bf 26}, 241--246 (1994).}
  
    \bibitem{gs}\newblock{T.~Schurmann and P.~Grassberger, {\em
      Chaos}, {\bf 6}, 414--427 (1996).}
    
    \bibitem{shannon}\newblock{C.~E.~Shannon, {\em Bell Sys.\ Tech.\ 
      J.} {\bf 27}, 379--423, 623--656 (1948).}
  
    \bibitem{note4}\newblock{Throughout this discussion we assume that
    the signal $x$ at one point in time is finite dimensional.  There
    are subtleties if we allow $x$ to represent the configuration of a
    spatially infinite system.}
  
    \bibitem{standish}\newblock{R.~K.~Standish, submitted to {\em
      Complexity International}. E-print: {\small \tt
      nlin.AO/0101006}.}
  
    \bibitem{crutchfield}\newblock{C.~R.~Shalizi and
    J.~P.~Crutchfield, to appear in {\em Journal of Statistical
      Physics} (2001). E-print: {\small \tt cond-mat/9907176}}.
  
    \bibitem{tpb}\newblock{N.~Tishby, F.~Pereira, and W.~Bialek, in
    {\em Proceedings of the 37th Annual Allerton Conference on
      Communication, Control and Computing}, B.~Hajek and
    R.~S.~Sreenivas, eds., University of Illinois, pp.~368--377
    (1999). E-print: {\small \tt physics/0004057.}}

\end{thebibliography}
\end{document}